\documentclass{webofc}
\usepackage[varg]{txfonts}   % Web of Conferences font
\usepackage{adsmacros}
\usepackage{booktabs}
\usepackage{graphicx}
\usepackage{natbib}
\usepackage[caption=false]{subfig}
\usepackage{xcolor}
\usepackage{siunitx}
\usepackage{hyperref}

\newcommand{\changed}[1]{\textcolor{black}{#1}}

\newcommand{\tup}[1]{\textup{#1}}
\newcommand{\tw}{\textwidth}

\newcommand{\msun}{$ \text{M}_\odot $}

\newcommand{\Rvir}{R_\tup{vir}}

\newcommand{\res}{\text{res}}
\defcitealias{bourdin:SZimaging}{B15}
\begin{document}
\title{Spectral imaging of X-COP galaxy clusters with the Sunyaev--Zel'dovich effect}
\author{\firstname{Anna Silvia} \lastname{Baldi}\inst{1,2}\fnsep
\thanks{\email{anna.silvia.baldi@roma2.infn.it}} \and 
	\firstname{Hervé} \lastname{Bourdin}\inst{1} 
	\and
	\firstname{Pasquale} \lastname{Mazzotta}\inst{1}
}

\institute{Universit\`{a} di Roma ``Tor Vergata'', Via della Ricerca Scientifica, I-00133 Roma, Italy
\and
           Sapienza Universit\`{a} di Roma, Piazzale Aldo Moro 5, I-00085 Roma, Italy}

\abstract{%
  The Sunyaev--Zel'dovich effect is the ideal probe for investigating the outskirts of galaxy clusters.
  To map this signal, we apply a spectral imaging technique which combines parametric component
  separation and sparse representations. Our procedure is an improved version of an existing algorithm, which now features 
  a better treatment of astrophysical contaminants, and the implementation of a new beam deconvolution.
  We use the most recent frequency maps delivered by \textsl{Planck}, and we consider the clusters analysed in the 
  \textsl{XMM} Cluster Outskirts Project (X-COP). In particular, we focus on the images of two clusters which may be
  possibly interacting with neighbouring objects, namely A2029 and RXCJ1825. We also highlight the advantages of the
  new beam deconvolution method, through a comparison with the original version of the imaging algorithm.
}
\maketitle
\section{Introduction}
\label{sec:intro}
Galaxy clusters are the largest virialized structures in the universe. As prescribed by the hierarchical scenario,
they form by accretion and merging of smaller collapsed objects. In the outermost cluster regions, at radial
distances from the cluster centre in the range $ R_{500} < r < 3 R_{200} \approx 3 \Rvir $\footnote{When the
subscript $ \Delta $ is used, it means that the volume of the sphere having radius $ R_{\Delta} $ encloses a density
which is $ \Delta $ times the critical density of the universe}, %~\cite{reiprich:clusteroutskirts},
gravity-driven accretion of matter towards the central halo is still ongoing.
A number of numerical works suggest that this process takes place along filaments and gaseous bridges connecting
the cluster to the cosmic web. Moreover, cluster outskirts are characterized by shocks and by an inhomogeneous,
clumpy distribution of the intra-cluster medium (ICM hereafter)~\cite[see e.g.][for a recent review]{walker:outskirtsreview}.
Such phenomena can induce non-negligible biases in the modelling of cluster thermodynamic 
properties and in the estimate of cluster masses, which are often based
on the assumption of hydrostatic equilibrium.

An optimal strategy to characterize cluster outskirts through observations is the combination of X-ray data
with measurements of the Sunyaev--Zel'dovich (SZ) effect~\cite{sz:70}. The latter is the Comptonization of cosmic
microwave background (CMB), which manifests as a distortion of its black body spectrum. The temperature
anisotropy produced by the SZ effect due to the random motion of the electrons (thermal SZ, tSZ hereafter) is
proportional to the integral of the thermal pressure of the gas along the observer's line of sight.
Thus, compared to X-ray observations, it is less sensitive to the decrease of the electron density at large cluster
radii, and it is an optimal probe of e.g. overpressure and shocks expected to occur in the outskirts.
In this context, the recent \textsl{XMM} Cluster Outskirts Project (X-COP)~\cite{xcop:presentation} exploits
the synergy between X-ray and tSZ data. This study unveiled the dynamical and thermodynamic properties of a
selected sample of clusters with an exquisite radial coverage~\cite[e.g.][]{xcop:thermoproperties}.

Thanks to its peculiar spectral signature, the tSZ effect can be reconstructed \changed{via component 
separation 
techniques
from multi-frequency data}.
To this end, a variety of methods relying both on non-parametric and parametric procedures
have been proposed.
In the recent work by~\cite{bourdin:SZimaging} (\citetalias{bourdin:SZimaging} hereafter), the authors introduce
a new algorithm for the spectral imaging of the tSZ effect, whose novelty consists of combining
the parametric approach with wavelet and curvelet decompositions.

In this work, we apply an improved version of the procedure by~\citetalias{bourdin:SZimaging} to the most recent
data released by the Planck Collaboration~\citep{planck:hfi2018}. Specifically, as discussed in~\cite{baldi:szimaging},
we included an optimized treatment
of astrophysical contaminants, namely thermal dust and bright point sources, and a new method to deconvolve \textsl{Planck}
beams. Focussing on the twelve massive and nearby clusters analysed in X-COP, we show that our method is suitable to
recover anisotropies located beyond $ R_{500} $ in their tSZ maps, with improved reliability and stability with
respect to the original implementation of the algorithm.

The paper is structured as follows. The X-COP data set is described in Sect.~\ref{sec:data}. Section~\ref{sec:method}
illustrates the tSZ imaging algorithm. The maps for the whole cluster sample are shown in Sect.~\ref{sec:results},
with a focus on two possibly interacting systems. We finally illustrate our conclusions and perspectives
in Sect.~\ref{sec:conclusions}.
\section{Data set}
\label{sec:data}
The target clusters of this work are the twelve massive objects ($ M_{200} > \num{6e14} $\msun), with redshift
in the range $ 0.04 < z < 0.10 $, considered in X-COP~\citep[see][for a detailed description of the data
set]{xcop:thermoproperties}.
These objects have been observed in the X-ray band by \textsl{XMM}-Newton,
and in the microwave band by \textsl{Planck}. The minimum signal-to-noise ratio (S/N 
hereafter) of \textsl{Planck} detection is greater than 12, and the angular size subtended
by the $ R_{500} $ radius is $ \theta_{500} > 10 $ arcmin for all the clusters, to ensure a good resolution
in the microwave band.

For our study, we used the most recent maps\footnote{Publicly available at \url{https://pla.esac.esa.int}} at the
six frequencies scanned by the High Frequency Instrument (HFI) onboard \textsl{Planck} satellite, namely 100, 143, 217,
353, 545 and 857 GHz. The beam FWHM varies across frequencies between $ \approx 10 $ arcmin and $ \approx 5 $ 
arcmin~\cite[see][]{planck:hfibeams2015}.
We extracted \changed{square patches centred at the location of the target clusters} in our sample. 
Such patches extend across $ \approx 4.3^{\circ} $ on each side, in order to cover sufficiently large radial
distances ($ \gtrsim 3\Rvir $ for all the clusters).
\section{Spectral imaging of the tSZ signal}
\label{sec:method}
Following~\citep{baldi:szimaging}, we modelled the temperature signal in each HFI map as the sum of the $ N_s $
sources to separate, $ s_i $, multiplied by their spectral energy density, $ f_i $. Each model map is therefore:
\begin{equation}
	\label{eqn:model}
	M(\nu, k; s) = \sum\limits_{i}^{N_s} f_i(\nu) \ s_i(k) + \eta(\nu,k) \ ,
\end{equation}
where $ \nu $ is the frequency, $ k $ is the index of a generic pixel, and $ \eta $ is the instrumental noise.
\changed{Considering the Galactic latitudes and the low redshift of our target clusters~\citep[see table 1
of][]{baldi:szimaging}, the three dominating physical components in HFI frequency range are}: the CMB, the tSZ
and Galactic thermal dust (td).
The tSZ is modelled in the non-relativistic limit~\citep[see][]{baldi:szimaging}, and the amplitude of
the tSZ signal is given by the Compton $ y $-parameter, defined as
\begin{equation}
	\label{eqn:y}
	y = \frac{\sigma_T}{m_e c^2} \int_\tup{los} p(l) \ dl \ ,
\end{equation}
where the integral is calculated along the line of sight (los), $ \sigma_T $ is the Thomson cross section,
$ m_e $ is the electron mass, $ c $ is the speed of light, and $ p(l) $ the thermal pressure of the ICM.
The spectral energy density of Galactic thermal dust is modelled as double grey body, i.e.
by assuming two populations of dust grains at different temperatures~\cite[see][for 
details]{baldi:szimaging}.

The parameters of the model, given by the source maps $ s_i $, are estimated via the minimization of a chi-square.
To take advantage of the efficiency of sparse representations in detecting localized and anisotropic features,
we minimized a weighted chi-square calculated by summing across all the $ N_{\nu} $ frequencies and the
$ N_\tup{pix} $ pixels:
\begin{equation}
	\label{eqn:chi2B15}
	\chi^2_{\Psi} = \sum_{\nu}^{N_{\nu}} \sum_{k}^{N_\tup{pix}} \
	\frac{\text{res}_{\Psi}^2(\nu,k;s)}{\sigma_\tup{HFI}^2(\nu,k)} \ ,
\end{equation}
where $ \sigma^2_\tup{HFI} $ is HFI variance, \changed{delivered with the intensity maps at each frequency},
and $ \res_{\Psi} $ is the wavelet transform of the residuals, being $ \Psi $ the wavelet basis function.
The residuals are calculated as the difference between the data maps, $ D_\tup{HFI} $, and the above model maps:
\begin{equation}
	\label{eqn:residuals}
	\text{res}(\nu,k;s) = D_\tup{HFI}(\nu,k)-M(\nu,k;s) \ .
\end{equation}
On the other hand, their wavelet transform is:
\begin{equation}
	\label{eqn:residuals15}
	\text{res}_{\Psi}(\nu,k;s) = \sum_n^{N_\tup{pix}} \bar a_{j_0,n}(\nu;s) \ \Phi_{j_0,n}(k) \ +
	\sum_{j=j_0}^{N_\tup{scales}} \sum_n^{N_\tup{pix}} a_{j,n}(\nu;s) \ \Psi_{j,n}(k) \ ,
\end{equation}
\changed{where $ n $ and $ j $ are the indices giving the translation and the scale of the wavelet, 
respectively}.
The first term in the right-hand side of Eqn.~\eqref{eqn:residuals15} represents the
approximation level, encoding information on the signal at the lowest spatial resolution,
$ 2^{-j_0} $. On the other hand, the second term contains information at high-resolution scales.
The total number of scales we used in the decomposition is $ N_\tup{scales} = 4 $, which is a suitable value for
our 256-pixel maps.
We chose the wavelet basis $ \Psi $ to be a B$_3 $ spline, being $ \Phi $ its dual function at the
largest scale, $ 2^{j_0} $.\\
To match the angular resolution of the data and model maps, we use a new deconvolution instead of the
iterative algorithm used in~\citetalias{bourdin:SZimaging}. Specifically, this procedure is applied directly
to the wavelet coefficients, $ \bar a_{j_0,n} $ and $ a_{j,n} $, of the approximation and detail levels, respectively.
Using the explicit expression of the residuals as in Eqn.~\eqref{eqn:residuals}, the wavelet coefficients are calculated
at each scale as
\begin{equation}
	\label{eqn:coeffresidualsdeconv}
	\begin{split}
		a_{j,n}(\nu;\Delta s) &= \sum_k^{N_\tup{pix}} \text{res}(\nu,k; \Delta s) \ \Psi^*_{j,n}(k) \\
		&=\sum_k^{N_\tup{pix}} \lbrace D_\tup{HFI} (\nu,k) - \mathcal{B}(\nu) \otimes [H k(\nu,k;\tilde s + \Delta s)+\\
		&+ (1-H) k(\nu,k;\tilde s - \Delta s)]\rbrace \ \Psi^*_{j,n}(k) \ ,
	\end{split}
\end{equation}
(the same holds for the approximation coefficients).
The model map is convolved with the instrumental beam, $ \mathcal{B} $, at each frequency. 
Denoting the mean value of the parameter as $ \tilde s $, its positive spatial variations,
$ \Delta s $, are coupled to the positive component of the wavelet by means of the Heaviside step function
$ H $. Vice-versa, negative variations, $ - \Delta s $, correlate with the negative component of the kernel via the
$ 1-H $ term.\\
To optimally map anisotropies in the tSZ and td signals, these sources are decomposed via a curvelet transform.
To retain only the relevant information in the signal, the coefficients are denoised by applying a soft
thresholding at $ 1\sigma $. The final component maps are then obtained by means of a suitable restoration operator, which
recombines the approximation and detail coefficients together.

To remove residual contamination from dust on large scales, we did not account for the approximation coefficients at
857 GHz when reconstructing the signal.
Indeed, at this frequency, which provides the highest resolution, the contribution from the signal at large
scales to the approximation coefficients is less significant. This procedure allowed us to remove gradients
affecting the tSZ signal of some clusters. %~\citep[see also fig. 1 of][]{hurier:a2319}.\\
Finally, we used the masks of bright point sources in the second \textsl{Planck} Catalogue of Compact
Sources~\cite{planck:compactsources2} to avoid the contamination from synchrotron and dusty galaxies which plagued
the reconstruction of the tSZ signal in a few cluster cases, namely  A3266, A85, and ZW1215.
\section{Results and discussion}
\label{sec:results}
\begin{figure}[tb]
	\centering
	\sidecaption
	\includegraphics[width=0.73\tw]{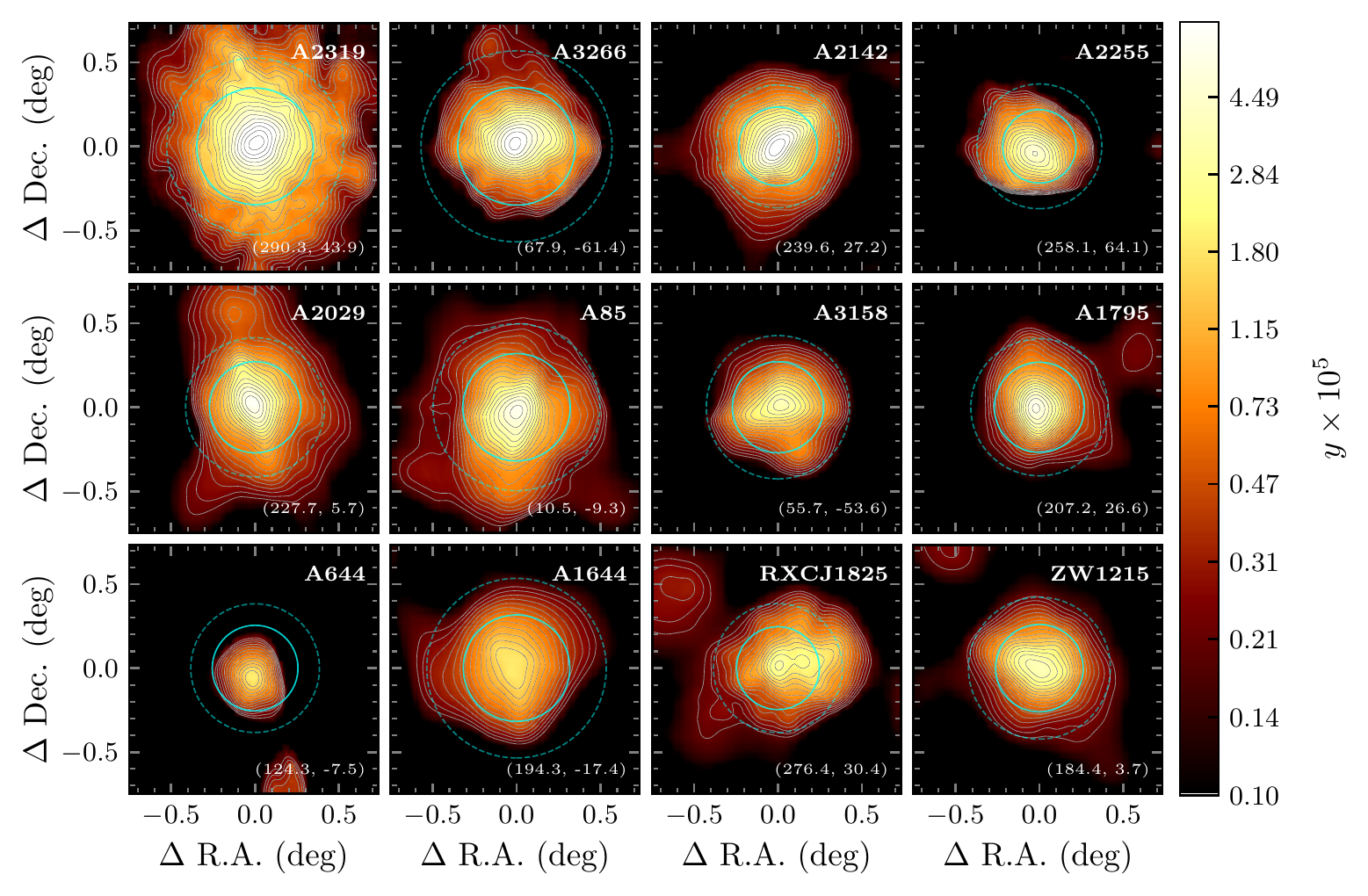}
	%HOME/newpimaging/curdecmaps
	\caption{Compton $ y $-parameter of X-COP clusters from our spectral imaging. %\num{4e-6} to \num{5e-5}.
		The solid and dashed cyan circles mark the $ R_{500} $ and $ R_{200} $ radii, respectively.
		The J2000 Equatorial coordinates of the cluster centres are reported in the bottom right corner of each map.}
	\label{fig:ymaps}
\end{figure}

The maps of the tSZ effect for the twelve X-COP clusters is shown in Fig.~\ref{fig:ymaps}.
It is possible to see a number of substructures and anisotropic features in the signal at radii beyond $ R_{500} $.
Moreover, thanks to the deconvolution, the ellipticity in the central regions of e.g. clusters A2142 and A1644 is
well recovered.\\
Among all the clusters, A2029 and RXCJ1825 are two particularly interesting cases. Indeed, both these objects have
been originally considered as possibly interacting systems with two neighbouring clusters~\cite{planck:interactingclusters}.
\begin{figure}[bt]
	\centering
	\subfloat[A2029] %
	{\includegraphics[height=0.32\tw]{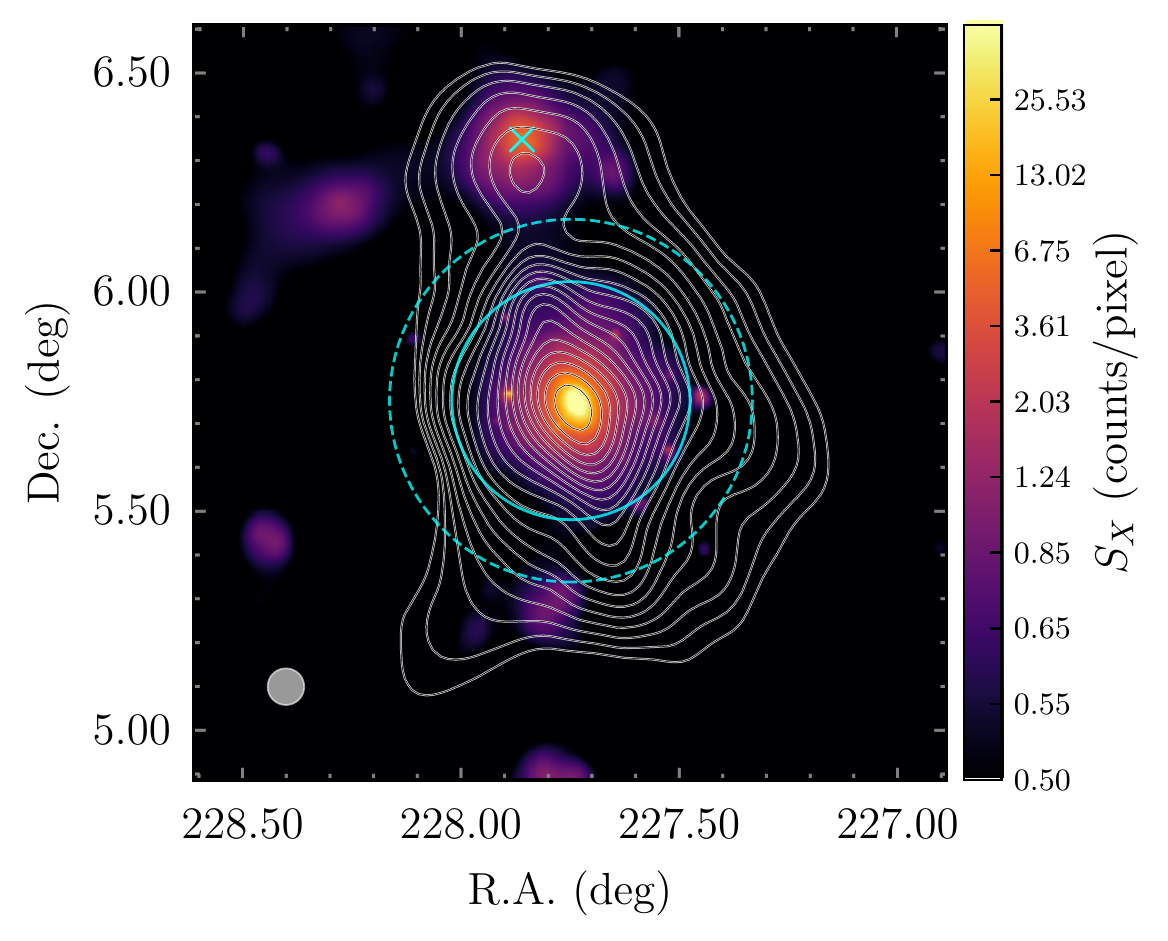} %
		\label{fig:a2029_XSZ}} %\quad
	\subfloat[RXCJ1825] %
	{\includegraphics[height=0.32\tw]{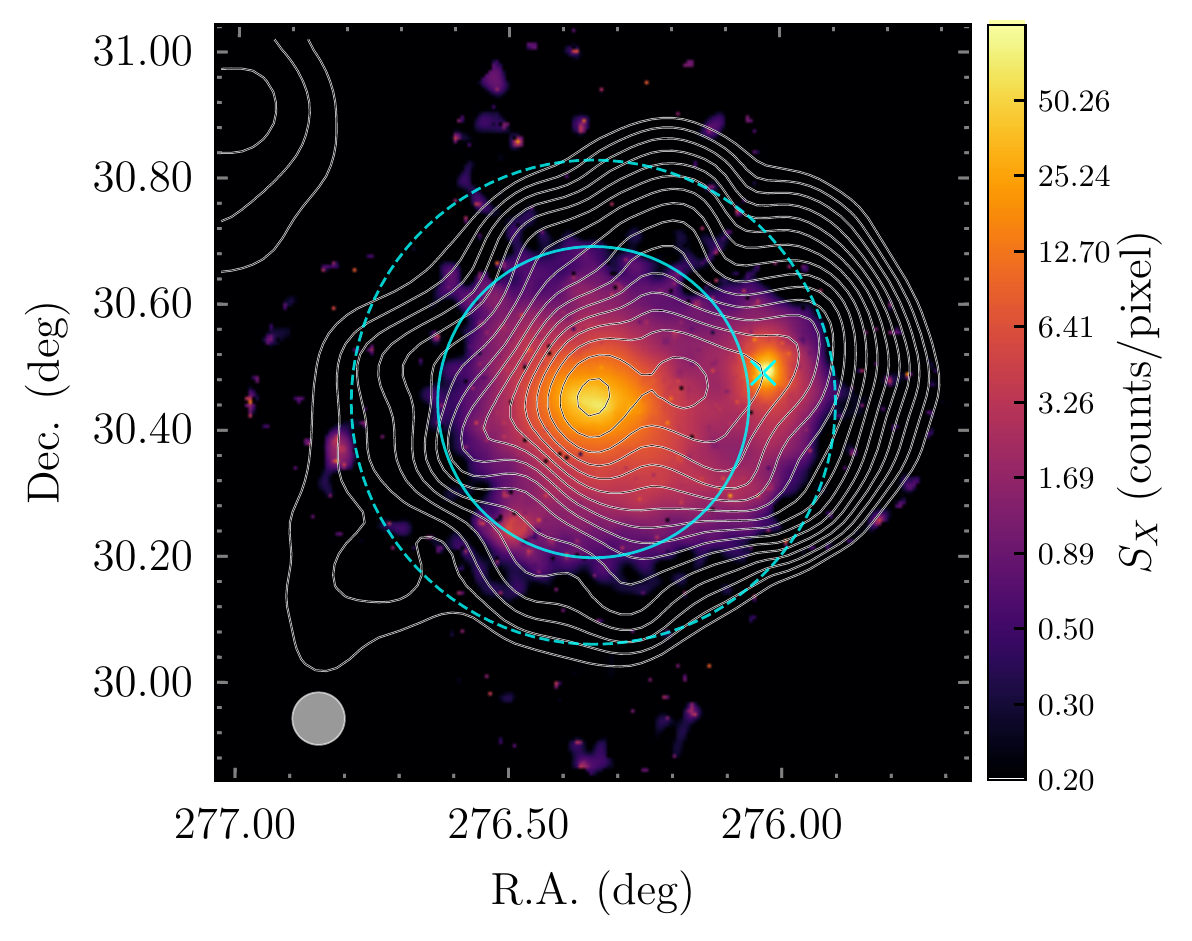}
		\label{fig:rxc1825_XSZ}}
	\caption{X-ray maps and tSZ contours of clusters A2029 and RXCJ1825 (see text for details).
	The cyan solid and dashed circles mark the $ R_{500} $ and $ R_{200} $ radii, respectively.
	The filled circles in the bottom left corner represent a 5 arcmin beam.}
	\label{fig:XSZ}
\end{figure}% SOURCE FILE: ~/pimaging_art/xray/XSZ.py
Figure~\ref{fig:XSZ} shows a comparison between the X-ray signal from A2029 and RXCJ1825, and the superimposed tSZ
contours from our spectral imaging algorithm. X-ray maps show the denoised, background-subtracted and vignetting-corrected
data from \textsl{ROSAT}/PSPC (for A2029) and from \textsl{XMM-Newton} (for RXCJ1825).
It can be seen that in both cases the large-scale features of the signal are similar in the two bands. We detect the
two neighbouring clusters A2033 and CIZA J1824.1+3029, whose positions are marked with cyan crosses in
Fig.~\ref{fig:XSZ}. In particular, A2033 is seen as a small blob in the tSZ signal \changed{detected with 
S/N $ \simeq 8 $}, located within the 5 arcmin beam with respect to its X-ray peak.
It is interesting to note that \changed{both these systems show significant elongations towards their
neighbouring clusters.}
These features could be useful in future investigations of the interaction between the objects in these
systems~\citep[see e.g.][]{planck:interactingclusters}.
The map of RXCJ1825 in Fig.~\ref{fig:rxc1825_XSZ} also shows a further elongated excess of signal towards the 
South-West region which has been confirmed also in recent X-ray and radio 
analyses~\cite{clavico:rxc1825}. This emission may be due to stripped gas from the interaction of the
central cluster with a nearby group of galaxies, which is now disrupted.

\begin{figure}[bth]
	\captionsetup[subfigure]{labelformat=empty}
	\centering
	% 0.185 \tht
	\subfloat %[Van Cittert deconvolution: map] %
	{\includegraphics[width=0.33\tw]{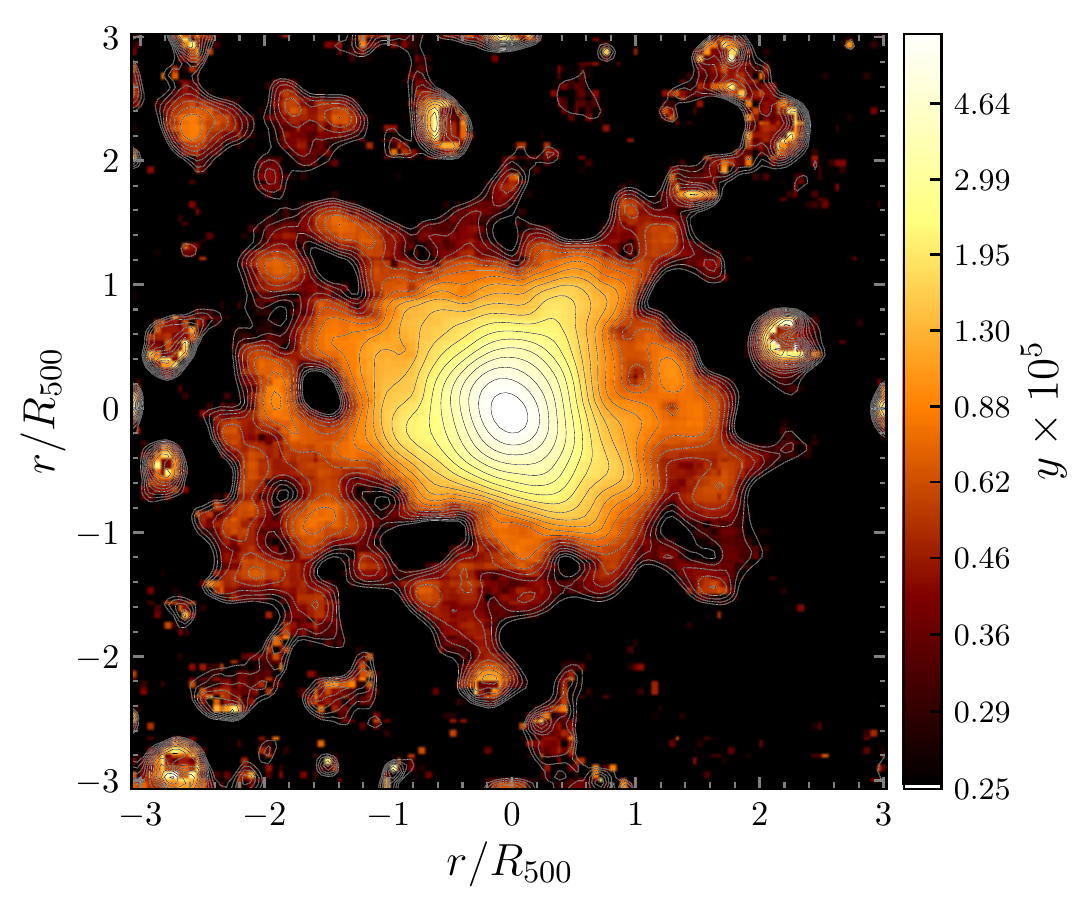}
		\label{fig:a2319old}}\quad
	\subfloat %[Van Cittert deconvolution: profiles] %
	{\raisebox{1.35ex}{%
	\includegraphics[width=0.33\tw]{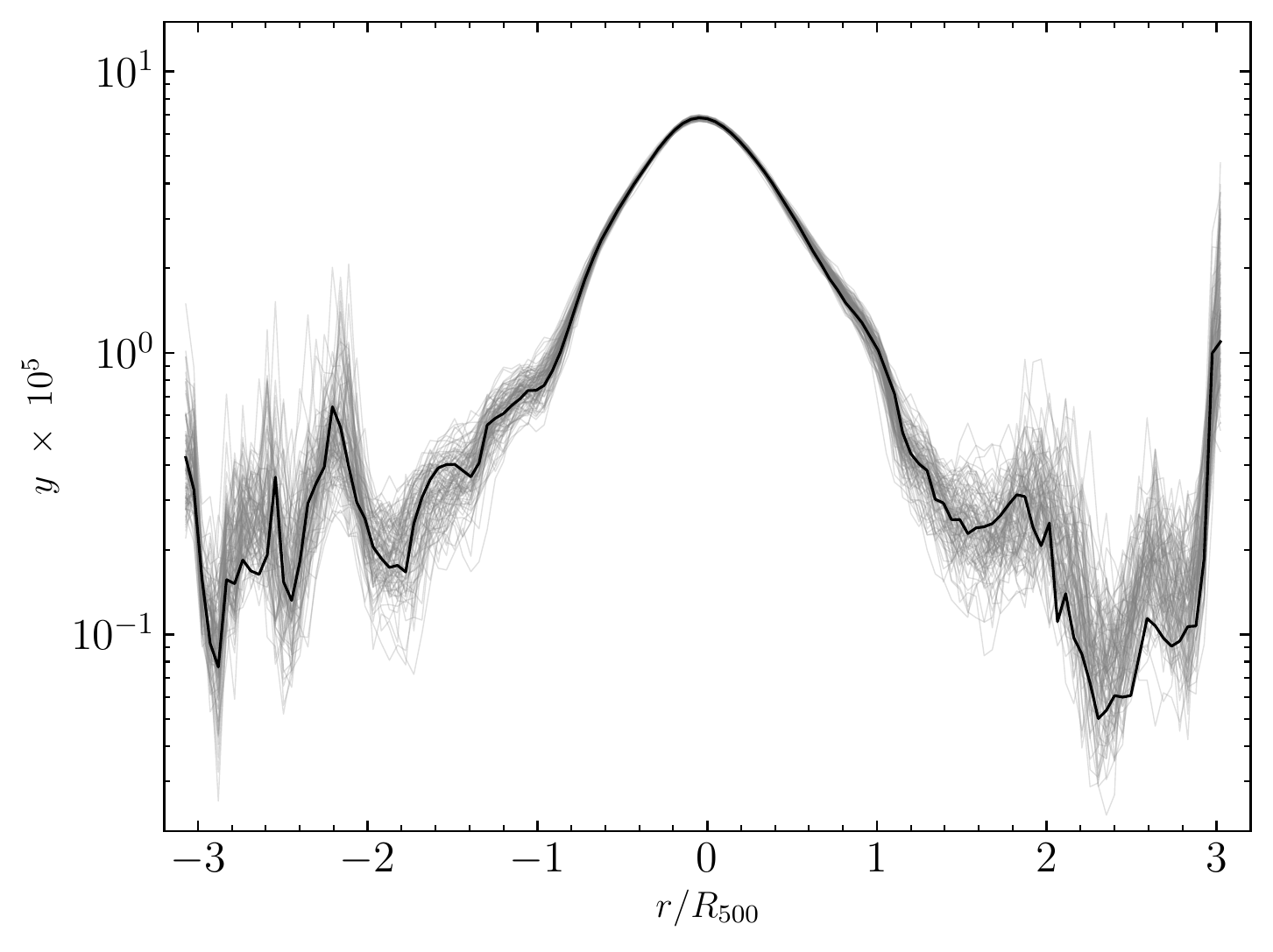}
		\label{fig:a2319boot_old_vert}}}\\
	\subfloat %[New deconvolution: map] %
	{\includegraphics[width=0.33\tw]{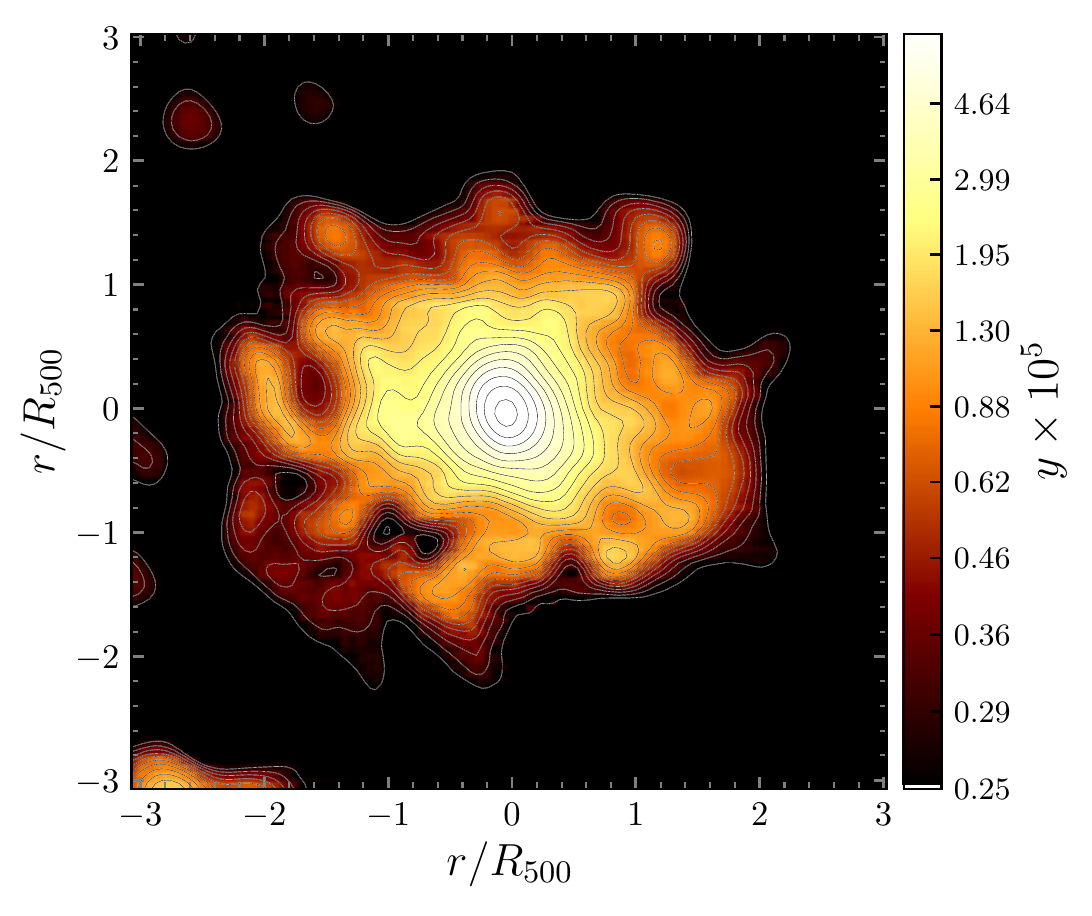}
		\label{fig:a2319new}}\quad
	\subfloat %[New deconvolution: profiles] %
	{\raisebox{1.35ex}{%
	\includegraphics[width=0.33\tw]{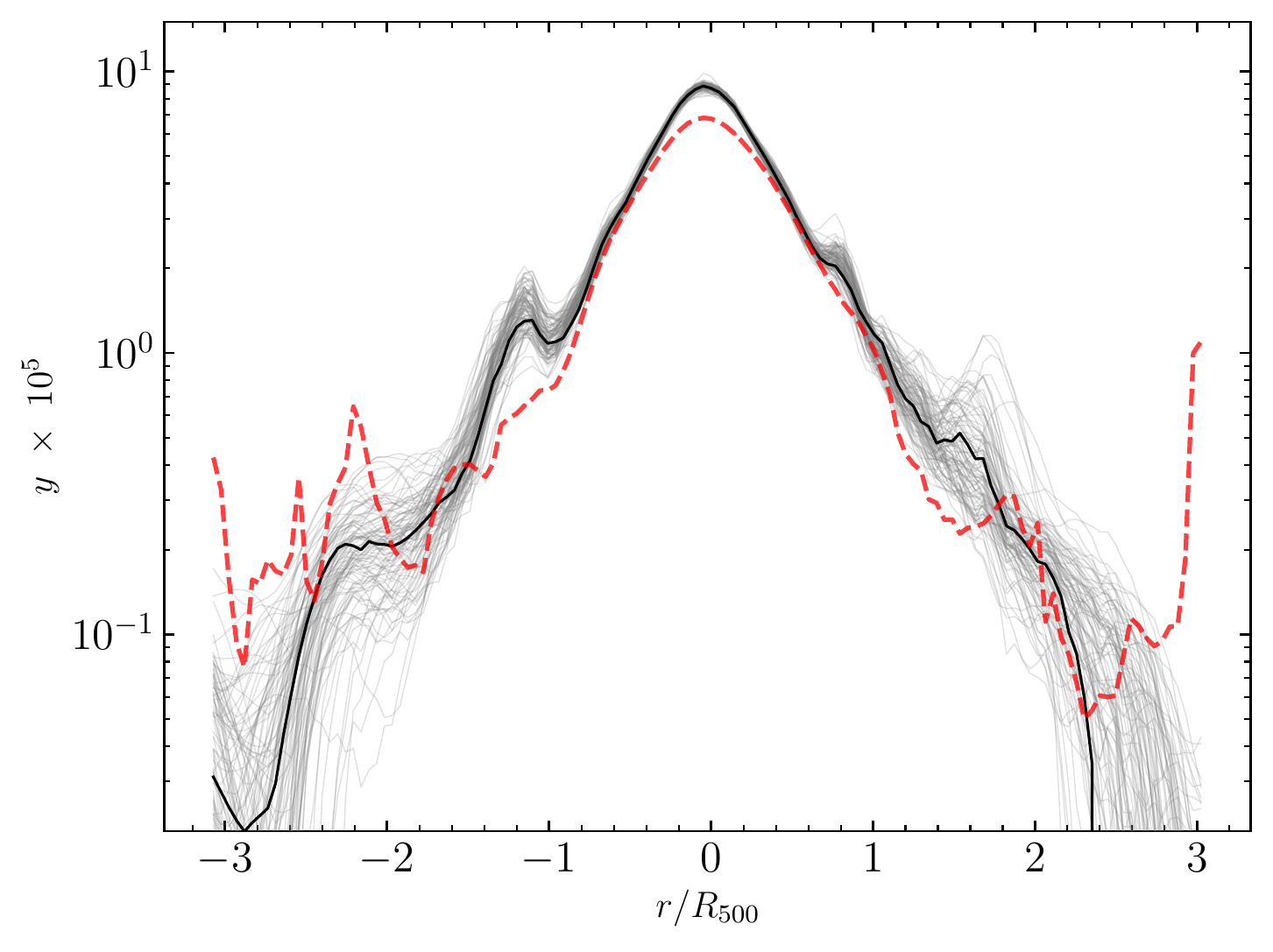}
		\label{fig:a2319boot_new_vert}}}
	%HOME/newpimaging/figures/cluster e HOME/pimaging_art/xray/XSZ2018.ipynb
	\caption{Compton $ y $-parameter of cluster A2319 reconstructed with~\citetalias{bourdin:SZimaging}
	version (top panels) and with the improved version presented in this work (bottom panels). See the full explanation in 
	the text.}
	\label{fig:a2319oldnew}
\end{figure}

To discuss the effects of the new deconvolution, we show in Fig.~\ref{fig:a2319oldnew} a comparison between the
tSZ signal of cluster A2319 obtained with the~\citetalias{bourdin:SZimaging} version of the algorithm (top panels),
and with the version presented in this work (bottom panels).
The left panels report the maps of the Compton $ y $-parameter. The right panels show as solid black lines the vertical
cuts passing through the centre of the corresponding maps, within a 10 arcmin-wide band.
The grey lines represent the same cuts extracted from \changed{100 different realizations of synthetic 
data},
as obtained with a bootstrap technique~\citep[see][for details]{baldi:szimaging}.
It can be seen that the new deconvolution allows a more stable reconstruction of the signal
in the outskirts, down to a few $ \num{e-6} $.
This is testified by the removal of the pixel-sized diverging artefacts which plague both the map and the profiles
in the top panels of Fig.~\ref{fig:a2319oldnew}, in regions corresponding to low-signal regimes.
Furthermore, the recovery of the value at the central peak is also improved, with a 23 per cent difference between the
two procedures. We highlight this finding in the bottom right panel of Fig.~\ref{fig:a2319oldnew}, where the profile
from the previous deconvolution is superimposed as a dashed red line to the profile from the new deconvolution (solid
black line).
%This fact is a consequence of the dependence of the peak value on the arbitrary number of iterations used to 
%adjust the resolution in the old deconvolution method employed in~\citetalias{bourdin:SZimaging}.
This fact is a consequence of the iterative nature of the deconvolution employed
in~\citetalias{bourdin:SZimaging}. Indeed, in this case the peak value increases with the arbitrary number of
iterations, which cannot be determined rigorously.
% SOURCE FILE: ~/newpimaging/scripts/compare_imaging.py
\section{Conclusions}
\label{sec:conclusions}
We presented an improved version of an existing spectral imaging algorithm to extract the tSZ signal
from multi-frequency maps.
Using the latest data by the \textsl{Planck} satellite, we produced the maps of the Compton
$ y $-parameter for the twelve clusters analysed in X-COP.
Our procedure, which is based on parametric component estimate and on sparse representations,
features a better treatment of thermal dust and bright point source contaminations. Moreover, it implements a new 
deconvolution
of \textsl{Planck} beams applied on the wavelet coefficients, yielding a more stable and reliable reconstruction
of the signal both in the cluster centre and in the outskirts.
Our results indicate we can detect a number of features, such as substructures and filaments, located beyond $ R_{500} $
in the majority of the objects. In particular, we report interesting results for A2029 and RXCJ1825, which
we compared with X-ray data.

A promising application of these improved tSZ images is the correction of radial pressure profiles by biases which
affect the outermost cluster regions. This topic has been already addressed in the recent literature (see the
contribution by F. Mayet in this proceeding volume). Our algorithm may be useful e.g. in combining results from large, nearby
systems detected by \textsl{Planck}, with the outcomes from the observations of high-redshift objects from competitive
instruments featuring a higher angular resolution, such as NIKA2~\citep{perotto:nika2performance}.
\bibliographystyle{woc}
\bibliography{BIBLIOGRAPHY}
\end{document}